\documentclass[twocolumn]{aastex6}

\usepackage[english]{babel}

\usepackage[T1]{fontenc}
\usepackage{ae,aecompl}

\usepackage{fancyhdr}
\usepackage{amsfonts}
\usepackage{amsmath}
\usepackage{amssymb}
\usepackage{layout}
\usepackage{graphicx}
\usepackage{multirow}
\usepackage{color}
\usepackage{multirow}

\usepackage{times}
\usepackage{natbib}
\def\be{\begin{equation}}
\def\ee{\end{equation}}

\newif\ifAMStwofonts
\AMStwofontstrue

\graphicspath{{./fig/}}

\shorttitle{model-independent approaches}
\shortauthors{Rezaei et al.}

\begin{document}

\title{Cosmographic parameters in model-independent approaches}
\author{Ahmad Mehrabi \altaffilmark{1}}
\author{Mehdi Rezaei \altaffilmark{1,2}}

\affil{\altaffilmark{1} Department of Physics, Bu-Ali Sina University, Hamedan, 65178, 016016, Iran }

\affil{ \altaffilmark{2}Iran meteorological organization, Hamedan Research Center for Applied Meteorology, Hamedan, Iran}


\begin{abstract}
Cosmographic approach, a Taylor expansion of the Hubble function, has been used as a model-independent method to investigate the evolution of the universe in the presence of cosmological data. Apart from possible technical problems like the radius of convergence, there is an ongoing debates about the tensions appear when one investigates some high redshift cosmological data. In this work, we consider two common data sets namely SNIa (Pantheon sample) and the Hubble data to investigate advantages and disadvantages of the cosmographic approach. To do this, we obtain the evolution of cosmographic functions using cosmographic method as well as two other well known model-independent approaches namely, the Gaussian process and the Genetic algorithm. We also assume $\Lambda$CDM model as concordance model to compare the results of mentioned approaches.  Our results indicate that the results of cosmography comparing with the other approaches, are not exact enough. Considering the Hubble data which is less certain, the results of $q_0$ and $j_0$ obtained in cosmography, provides a tension at more than $3\sigma$ away from the best result of $\Lambda$CDM. Assuming both of data samples in different approaches we show that the cosmographic approach, because of providing some biased results, is not the best approach for reconstruction of cosmographic functions, especially at higher redshifts.

\end{abstract}


\section{Introduction}
The $\Lambda$CDM model as the simplest explanation for the accelerated expansion of our universe, was built assuming cold dark matter and a cosmological constant $\Lambda$ in a Friedman geometry. It has proven so far to be the most successful cosmological model which accounts for the dynamics and the large-scale structure of our universe. This scenario matches well with the main cosmological observations including SNIa \cite{SupernovaSearchTeam:1998fmf,SupernovaCosmologyProject:1998vns,SupernovaCosmologyProject:2008ojh,Scolnic_2018}, baryons acoustic oscillations (BAO)\cite{Tegmark2004a,Cole:2005sx,Percival:2009xn,Reid:2012sw,Alam:2016hwk} and cosmic microwave background (CMB)\cite{Komatsu2009,Komatsu2011,Planck:2015bue,Aghanim:2018eyx}.

 Nevertheless, besides this fascinating advantages, $\Lambda$ cosmology suffers from persistent tensions of various degrees of significance between some of different observational data. One of the most intriguing tensions is the significant deficiency in the current value of Hubble parameter, $H_0$,predicted by the Planck team \cite{Aghanim:2018eyx} using the base $\Lambda$CDM model when compared with the values predicted by direct local measurements in \cite{Riess:2018byc,Riess:2019cxk,Freedman:2019jwv}. The other 
  tension concerns the discrepancy between the amplitude of matter fluctuations from large scale structure data from 
  \cite{Macaulay:2013swa} and its value that predicted by $\Lambda$ based on the CMB experiments. Moreover the discrepancy has been confirmed in \cite{Nunes:2021ipq} by comparing the value of matter fluctuations amplitude from a data combination of BAO+RSD+SNIa and the results of the Planck team. 
Furthermore, the BAO results from the Lyman-$\alpha$ forest measurement, reported in \cite{BOSS:2014hwf} predicts a smaller value for matter density parameter  comparing with the value predicted by  CMB data. In addition, the $\Lambda$CDM suffers from some serious theoretical problems of fine tuning and cosmic coincidence \cite{Weinberg1989,Padmanabhan2002,Copeland:2006wr}.
  
   To resolve these theoretical and observational problems, the underlying assumptions of $\Lambda$ cosmology should be examined. Two of these key assumptions are: $i)$ Dark energy is non-evolving and $ii)$ General relativity is applicable at all scales. Cosmologists have proposed different alternatives to these assumptions, coming in the form of dynamical dark energy models and modification of the gravity \cite[see also][]{Veneziano1979,Erickson:2001bq,Thomas2002,Caldwell2002,Padmanabhan2002,Gasperini2002,Sotiriou:2008rp,Myrzakulov:2012qp,Gomez-Valent:2014fda}. Despite various investigations which have been developed in the theoretical part, much of this models has been very well fitted by observational data sets \cite[for a review see][]{Malekjani:2016edh,Rezaei:2017yyj, Rezaei:2017hon,Malekjani:2018qcz,Lusso:2019akb,Rezaei:2019roe,Rezaei:2019hvb,Lin:2019htv,Rezaei:2019xwo,Noller:2020afd,Rezaei:2020mrj,Rezaei:2021qpq}. However, the nature of dark energy is still unknown. The evolution of the universe can be quantitatively investigated through different cosmological observations. It can be very helpful to obtain information on the universe directly from cosmological observations without assuming any hypotheses for dynamic of the universe. 
For extrication from dependency on a cosmological model, various model-independent approaches have been proposed. 
The artificial neural network (ANN) is a model-independent approach which based on machine learning and widely used in regression and estimation tasks \cite{Cybenko}. Recently, methods based on ANNs have shown outstanding performance in solving cosmological problems in both accuracy and efficiency. For example in \cite{Wang:2019vxv}, authors examined the ANN method by reconstructing functions of the distance-redshift relation of SNIa and the Hubble parameter $H(z)$. Furthermore, they estimated cosmological parameters using the reconstructed functions and found that their results are consistent with those obtained directly from observational data sets. Another well known model-independent approach which is commonly used in the literature for testing the fitting capability of cosmological models with observations, is cosmography, proposed in \cite{Alam:2003sc,Sahni:2002fz}. 
In \cite{RodriguesFilho:2017pjx}, authors considered the cosmographic approach and expanded the luminosity distance as well as the Hubble parameter around an arbitrary scale factor to estimate the cosmographic parameters, $H_0, q_0, j_0$ and $s_0$. In addition, the authors of \cite{Rezaei_2020} used model-independent cosmographic approach to compare $\Lambda$CDM with three different DE parameterizations. Using a Markov Chain Monte Carlo analysis, they put constraints on different cosmographic parameters. Comparing the results with those obtained from different DE scenarios, they showed that the concordance $\Lambda$CDM has a serious tension with high redshift data samples. Moreover, the authors of \cite{Capozziello:2011tj} investigated the possibility to extract the model-independent information of the dynamics of the universe assuming cosmographic approach. Their results indicated a considerably deviation from the $\Lambda$CDM model. Upon cosmographic approach, authors of \cite{Capozziello:2018jya} constrained the late time evolution of cosmic fluid using the low-redshift data sets coming from SNIa, BAO, $H(z)$, $H_0$ , strong-lensing time-delay as well as the Mega-maser observations for angular diameter distances. 

The other model-independent method is the Gaussian Process (GP). The method is a generalization of the Gaussian random variables over a function space and can describe a data set in a model-independent manner \cite{GP_book,Seikel:2012uu}. Gaussian Process as a powerful nonlinear interpolating tool and is widely used in cosmology\cite{Seikel:2012uu,Shafieloo_2012,Liao_2019,Wang:2019ufm,Gomez-Valent:2018hwc,Liao:2019qoc,Mehrabi_2020}. The authors of \cite{Bonilla:2020wbn} considered some cosmological data and applied the GP method to perform a joint analysis and put constraints on $H_0$ and some properties of DE like the equation of state, the sound speed of DE perturbations as well as the ratio of DE density evolution. 


Along with the GP, we consider another model-independent method namely the Genetic Algorithm (GA). In this approach, an arbitrary reconstruction can be built from some given basic functions using genetic evolutionary process. In fact a group of function sets have been evolved through crossover and mutation operation to find a function closer to a given data points. This method has been used to investigate some cosmological data in \cite{Bogdanos_2009,Nesseris_2010,Nesseris_2012,Arjona:2019fwb,Arjona_2020,arjona2021novel}

 Our paper is organized as follows: In Sec.\ref{sec:csomo_lcdm}, we describe the cosmography in a model dependent cosmology as well as the $\Lambda$CDM. In Sec. \ref{sec:model_ind}, we introduce cosmographic approach and present how the coefficient of the Taylor expansion related to the cosmographic parameters at present time. In addition, details of two other model-independent methods are presented in the section. Moreover, the results and discussions have been given in Sec. \ref{sect:res} and finally in Sec. \ref{sec:con} we summarize the main aspects of the results and conclude.

\section{Cosmography parameters in a cosmological model}\label{sec:csomo_lcdm}

 In general relativity, the evolution of space-time and included components are given by the Einstein field equation. Given a geometry, one can compute the evolution of each component in the Universe. Assuming a flat FRW geometry, the Hubble parameter describes how our universe evolve through cosmic time or redshift. The Hubble parameter in a universe containing matter, radiation and a dark energy with EoS $w(z)$ is given by:  
\begin{eqnarray}\label{wcd}
H^2(z)=H^2_0[\Omega_{m0}(1+z)^{3}+\Omega_{r0}(1+z)^4\\ \nonumber
+(1-\Omega_{m0}-\Omega_{r0})e^{3\int \frac{dz}{1+z}(1+w(z)) }]\;,
\end{eqnarray}

where $H_0$ and $\Omega_{m0}$( $\Omega_{r0}$) are the present value of the Hubble parameter and pressure-less matter energy density (radiation energy density) respectively. In the case of $\Lambda$ cosmology we have $w(z)=-1$ and so:
\begin{eqnarray}\label{lam}
H_{\Lambda}^2(z)=H^2_0(\Omega_{m0}(1+z)^{3}+\Omega_{r0}(1+z)^4+(1-\Omega_{m0}-\Omega_{r0}))\;,
\end{eqnarray}
Considering recent time evolution, the $\Omega_{r0}$ is quit smaller than other terms and can be omitted safely. 
The current rate of the expansion $H_0$ is very important quantity in understanding our universe. In addition to the rate of the expansion, other quantities have been defined considering derivative of this function.
\begin{eqnarray}\label{eq:q-j-h1}
&q(z)& = (1+z)\frac{H'(z)}{H(z)} - 1,\\
&j(z)& = (1+z)^2[\frac{H''(z)}{H(z)}+(\frac{H'(z)}{H(z)})^2] - 2(1+z)\frac{H'(z)}{H(z)}  + 1 \label{eq:q-j-h2}
\end{eqnarray}

These functions are the deceleration and jerk parameters respectively. The $q(z)$ and $j(z)$ are two important quantities in the context of the cosmographic analysis.

In the $\Lambda$CDM model, the cosmographic parameters are  
\begin{eqnarray}\label{qwcd}
&q(z)&=\frac{\Omega_{m0}(1+z)^{3}-2(1-\Omega_{m0})}{2 E^2(z)}\\ \nonumber
&j(z)&=1\;,
\end{eqnarray}
where $E(z)=H(z)/H_0$ is the dimensionless Hubble function. The deceleration parameter depends on the matter density but the jerk parameter is a constant. Any deviation from these values could be considered as a tension/discrepancy in the concordance $\Lambda$CDM. So computing these quantities directly from observation is quit useful in understanding underlying model of our universe. 

In this work, we use the following cosmological data to study the cosmographic parameters.
\begin{itemize}
	\item The most updated and precise measurement of luminosity distance at present time, the Pantheon sample \cite{Scolnic_2018}. This sample contains 1048 spectroscopically confirmed SNIa in the redshift range $0.01<z<2.26$. Note that, to prevent the degeneracy between $H_0$ and the absolute magnitude of the SNIa, we set $M_B=-19.3$ throughout our analysis.

	\item The Hubble parameter data which includes measurement of the cosmic chronometer as well as the Radial BAO from \cite{Farooq:2016zwm}. In addition to these 38 data points, we consider $H_0$ measurement from nearby SNs \cite{Riess:2019cxk}. Note that the BAO data points are correlated but for the sake of simplicity, we ignore these correlations in our analysis. 
\end{itemize} 

In the context of the $\Lambda$CDM, we perform a Bayesian inference to place constraints on the model parameters and then find cosmographic parameters in the model. Performing a Markov chain Monte Carlo (MCMC), the best value of parameters are $\Omega_{m0}=0.285\pm 0.012$ and $H_0=71.84\pm0.22$ for the SNIa data and $\Omega_{m0}=0.239\pm 0.015$ and $H_0=72.1\pm1.1$ for the Hubble data. It is then straightforward to compute the best fit cosmography as a function of redshifts as well as the $95\%$ $(1.96\sigma)$ confidence intervals. The results will be shown along with the results of other methods later in Sec. \ref{sect:res}. 

In order to use the SNIa data to constrain the $H_0$ parameter, the value of absolute magnitude should be fixed. In fact, the theoretical apparent magnitude of a standard candle is given by $m = 5\log_{10}D_L(z)+M_B+25$, where $D_L(z)$ is the luminosity distance. Considering $\Lambda$ cosmology, the apparent magnitude is given by:

\begin{equation}
m = 5\log_{10}c(1+z)f(\Omega_{m0})-\log_{10}H_0+M_B+25,
\end{equation}

where we have:
\begin{equation}
 f(\Omega_{m0}) = \int_0^z \frac{dz}{\sqrt{\Omega_{m0}(1+z)^3+(1-\Omega_{m0})}}.
 \end{equation}
 According to the above formula, given a sample of SNIa, one can constrain the value of $\Omega_{m0}$ as well as constant $a=-\log_{10}H_0+M_B+25$. Hence, by setting a value for $M_B$, the $H_0$ will be constrained uniquely. Since there is a degeneracy in this case, sometimes the $H_0$ tension is assumed as a tension on the $M_B$ parameter \cite[see also][]{Camarena:2021jlr,Efstathiou:2021ocp,Nunes:2021zzi}. Moreover, we rerun our code assuming a Gaussian prior on $M_B$ to show how different values of $M_B$ affect the estimation of $H_0$. To this aim, we assume two different value of $M_B = -19.401\pm0.027$ and $M_B = -19.244\pm0.037$ from \cite{Camarena:2021jlr,Efstathiou:2021ocp,Nunes:2021zzi} and obtain $H_0=68.67\pm0.82$ and $H_0=74.23\pm1.41$ respectively.
Notice that, in this work, our main objective is estimation of cosmographic parameters in some model-independent methods. To do this, we have to fix a value for $M_B$ to have a unique value of $H_0$ and assuming another value for $M_B$, the results will be shifted accordingly.

\section{Cosmography in a model-independent method}\label{sec:model_ind}
Nowadays, many different cosmological models have been trying to explain the accelerated expansion of the universe. All of these scenarios depend on the theory of gravitation and virtues of included contents. Since a model might bias and gives an incorrect value for a quantity, it will be helpful if we have an approach to reconstruct cosmological quantities directly from an observation and without assuming any model. In this section we will briefly introduce three of the most well-known model-independent methods and use them to study the evolution of universe in terms of cosmographic functions. The first one comes from a Taylor expansion in the Hubble parameter and other two approaches are GA and GP.
  
\subsection{Cosmography in the Taylor expansion of the Hubble parameter }\label{sec:csomo_exp}
In this part a Taylor expansion of the Hubble parameter has been used to reconstruct the cosmographic parameters. This method is usually called cosmographic approach and has been used in several works in cosmology \cite{Capozziello:2017ddd,Capozziello:2019cav,Li:2019qic,Mandal:2020buf,Capozziello:2021xjw}. This series up to the forth order in redshift $z$ around $z = 0$ has the following form:

\begin{eqnarray}\label{hexpandz}
H(z)=H\vert_{z=0}+\frac{dH}{dz}\vert_{z=0} \frac{z}{1!}+\frac{d^2H}{dz^2}\vert_{z=0} \frac{z^2}{2!} + \nonumber\\
\frac{d^3H}{dz^3}\vert_{z=0} \frac{z^3}{3!}+\frac{d^4H}{dz^4}\vert_{z=0} \frac{z^4}{4!}\;,
\end{eqnarray}
It is worth noting that the above series expansion implies fundamental difficulties with the convergence and the truncation of the series. Thus the function should be used only in its radius of convergence region which is $(z<1)$. In order to overcome this problem, following  \cite{Capozziello:2011tj} we use the y-redshift $y=\frac{z}{z+1}$, an improved redshift definition that commonly used in literature. Assuming y-redshift, we can improve the convergence domain. ( to see more detail refer to \cite{Capozziello:2011tj,Li:2019qic,Capozziello:2020ctn}). Applying improved redshift instead of $z$, Eq. \ref{hexpandz} takes the following form:
\begin{eqnarray}\label{hexpandy}
H(y)=H\vert_{y=0}+\frac{dH}{dy}\vert_{y=0} \frac{y}{1!}+\frac{d^2H}{dy^2}\vert_{y=0} \frac{y^2}{2!} + \nonumber\\
\frac{d^3H}{dy^3}\vert_{y=0} \frac{y^3}{3!}+\frac{d^4H}{dy^4}\vert_{y=0} \frac{y^4}{4!}\;.
\end{eqnarray}
Assuming the relations of cosmographic parameters (the Hubble, deceleration, jerk, snap and the lerk ), the cosmographic parameters ($H_0,q_0, j_0, s_0 , l_0$) and changing the time derivatives in to $y$ we will have (for details of the mathematical relations we refer the reader to \citep{Rezaei_2020}):
\begin{eqnarray}\label{ey}
H(y)=H\vert_{y=0}(1+k_1 y+\frac{k_2y^2}{2}+\frac{k_3y^3}{6}+\frac{k_4y^4}{24})\;.
\end{eqnarray}
where different $k_i$ are:

\begin{eqnarray}\label{k1}
k_1=1+q_0\;,
\end{eqnarray}

\begin{eqnarray}\label{k2}
k_2=2-q^2_0+2q_0+j_0\;,
\end{eqnarray}
\begin{eqnarray}\label{k3}
k_3=6+3q^3_0-3q^2_0+6q_0-4q_0j_0+3j_0-s_0\;,
\end{eqnarray}

\begin{eqnarray}\label{k4}
k_4=-15q^4_0+12q^3_0+25q^2_0j_0+7q_0s_0-4j^2_0-\nonumber \\
16q_0j_0-12q^2_0+l_0-4s_0+12j_0+24q_0+24\;.~~
\end{eqnarray}

\begin{figure*}[h]
	\centering
	\includegraphics[width=12 cm]{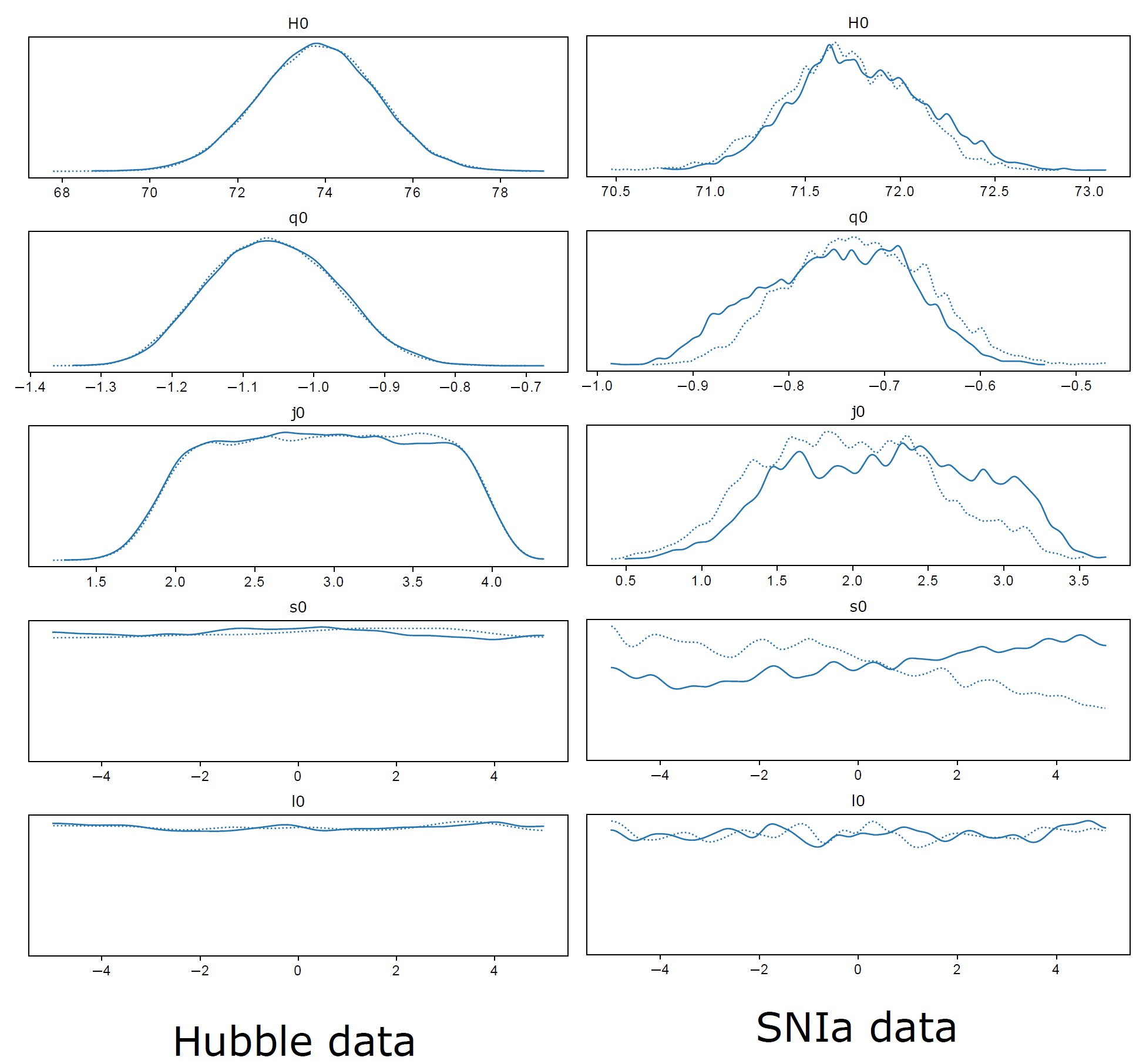}	
	\caption{The trace of the free parameters in the cosmographic approach. The left panel: trace of parameters using the Hubble data. The right panel: trace of parameters using  SNIa data.}
	\label{fig:trace}
\end{figure*}

Now having the function of Hubble parameter $H(y)$,  we can put constraints on cosmographic parameters using different observational data sets. To this aim, we repeat the procedure which we performed in the previous section and find the cosmographic parameters in this approach. Considering a wide flat prior for the free parameters and performing a Bayesian inference using PyMC3 package \cite{Salvatier2016}, we find the trace of parameters as it is shown in Fig.(\ref{fig:trace}). For both data sets, prior and posterior of $l_0$ and $s_0$ are the same and so data can't constrain these parameters at all. Assuming the best value of parameters, we have the best Hubble parameter from Eq. (\ref{ey}) as well as the best cosmographic parameters by substituting the Hubble parameter in Eqs. (\ref{eq:q-j-h1}),(\ref{eq:q-j-h2}). 

\subsection{Gaussian process}   
GP approach is a sequence of Gaussian random variables (RV), which can be modeled by considering a multivariate Gaussian distribution. In this case, the diagonal (off-diagonal) terms in the covariance matrix give uncertainty at each point (correlation between different points). Thus we can model a data set using a Gaussian process as $$f(x)\sim GP(\mu(x),K(x,\tilde{x}))$$ where $K(x,\tilde{x})$ is the kernel function and $x$ are the observational points. The $\mu(x)$ provides the mean of the RV at each $x$. Given a set of observation $(x,y)$ and a kernel function, it is straightforward to find ($y^{\star}$) at a set of arbitrary points $(x^{\star})$. To do this, first of all the mean and covariance matrix of a multivariate Gaussian distribution should be computed from \cite{10.5555/1162254}
\begin{eqnarray}\label{eq:GP}
\mu^{\star} &=& K(x,x^{\star})[K(x,x^{\star})+C_D]^{-1}Y\\
\Sigma^{\star} &=& K(x^{\star},x^{\star}) - K(x^{\star},x)[K(x,x^{\star})+C_D]^{-1}K(x,x^{\star}),
\end{eqnarray}
and then the $y^{\star}$ can be obtained from sampling of the distribution. In above equations, $C_D$ and $Y$ are the covariance and column vector of data points respectively. 

The most well-known kernel function is the squared exponential which is given by $$K(x,\tilde{x}) = \sigma_f^2\exp{\frac{-(x-\tilde{x})^2}{2\sigma_l^2}}$$
 In this case, $\sigma_f^2$ and $\sigma_l$ are two hyper-parameters which must be constrained using a Bayesian inference \cite{GP_book,Seikel:2012uu}. By employing the squared exponential kernel in the {\it{scikit-learn}} library \cite{scikit-learn}, we reconstruct the Hubble parameter (luminosity distance) in the case of the Hubble (SNIa) data sets.    

Having obtained the luminosity distance, it is straightforward to compute the Hubble parameter. The quantity is related to the comoving distance via
\begin{equation}\label{eq:lum-dis1}
D_L(z) = (1+z)D(z),
\end{equation}
and the comoving distance $D(z)$ in a flat geometry is 
\begin{equation}\label{eq:com}
D(z) = \int_0^z \frac{dx}{H(x)},
\end{equation}
so in this case, we need to take a derivative to obtain the Hubble parameter,
\begin{equation} \label{eq:H_from_com}
H(z) = \left[D'(z)\right]^{-1}.
\end{equation}  

As we mentioned before, we need to take derivative to compute the cosmographic parameters from the Hubble parameter. In the GP scenarios, it is an easy task to compute not only the first derivative but also other high order derivatives. In fact, derivative of a GP is another GP and one should only compute the mean and covariance matrix of the derivative. All related formula have been given in  \cite{GP_book,Seikel:2012uu}.  

Furthermore, we know that the kernel function might influence the results of GP. Thus, considering the standard Gaussian Squared-Exponential as the main kernel in our analysis, we have used two other Matern class of kernels, Matern $(\nu=7/2)$ (M72) and Matern $(\nu=9/2)$ (M92) (see \cite{Mehrabi_2020} for more details on these kernels). The results for both the SNIa and Hubble data have been presented in Tab.(\ref{tab:kernel}). As it is clear from the results, all values are quite in agreement at $1\sigma$ level.

\begin{table*}
	\centering
	\begin{tabular}{|r  c  c  c  c c  c  c c |}
		\hline \hline
		&        &   Hubble data    &   &$\vert$&   & SNIa data &   \\
		\hline
		Kernel $\vert$ & Gaussian kernel & M72 kernel & M92 kernel &$\vert$& Gaussian kernel & M72 kernel & M92 kernel\\
		\hline 
		$H_0$$\vert$ & $73.44\pm1.40$ & $73.48\pm1.36$ & $73.45\pm 1.36$ &$\vert$& $71.92\pm0.38$ & $71.86\pm 0.40$ & $71.99 \pm 0.45$\\
		\hline
		$ q_0 $$\vert$ & $-0.856\pm0.111$& $-0.866\pm0.129$ & $-0.862\pm 0.118$ &$\vert$& $-0.558\pm0.040$ & $-0.549 \pm 0.044$ & $-0.565 \pm 0.056$\\
		\hline
		$ j_0 $$\vert$  &$1.30\pm0.37$& $1.29\pm 0.47$ & $1.30\pm 0.42$ &$\vert$& $0.85\pm0.12$ & $0.819\pm 0.165$ & $0.948\pm 0.220$\\
		\hline \hline
		\end{tabular}
	\caption{Comparison between different cosmographic parameters and their $1\sigma$ uncertainties which are obtained using different kernel functions in GP approach.}\label{tab:kernel}
\end{table*}

 \subsection{Genetic algorithm}
The latest method which we have been used in our analysis is the GA. The early idea of the method is come from Darwin's theory of evolution. According to this scenario, over time and generations, different species evolve and become more compatible with the nature.
The GA can be regarded as a simulation of the above evolutionary process that based upon an arbitrary set of functions.  In this process, a set of trial functions evolves as time passes by through the effect of the stochastic operators of crossover, i.e. the joining of two or more candidate functions to form another one, and mutation, i.e. a random alteration of a candidate function. This process is repeated several times using different random seeds to explore whole functional space and ensure the convergence. Since the GA is constructed as an accidental approach,the probability that a set of functions will bring about offspring is principally assumed to be proportional to a fitness function. We use the  $\chi^2$ as the fitness function which gives the information on how good every individual agrees with the data.

One of the well known methods of doing genetic programming is the symbolic regression. Given a set of functions, the symbolic regression generates many mathematical expressions to describe a data set. Each of these expressions has their own $\chi^2$ value which decreases after each generation. Contrary to GP, the GA is independent of any kernel function which might affects the results. In this work, we use the public package {\it{gplearn}} which is an extension of {\it{scikit-learn}} machine learning library to perform a symbolic regression. We examine different set of basic functions and input parameters and find out that simple $(+,-,\times)$ function set, population-size = 2000, tournament-size = 30, p-crossover=0.9, p-hoist-mutation = 0.03 and p-point-mutation = 0.03 are good inputs to produce reconstructions efficiently.  

We run our code several times with different random seeds to generate many reconstructions and then select those with $\chi^2$ in the range of $(\chi^2_{min},\chi^2_{max})$ where these quantity can be calculated from the $\chi^2(k)$ distribution functions. In fact, these upper and lower band can be calculated by fixing amount of probability in the $\chi^2(k)$ distribution. For each data set, we compute $k$, the number of degree of the freedom and then from the $\chi^2(k)$ distribution function, $(\chi^2_{min},\chi^2_{max})$ have been computed by setting p-value =0.05. After having a sample of reconstructions, a similar to GP procedure has been done to compute the mean and standard deviation at each redshifts. 



\section{Results and discussions}\label{sect:res}
\begin{figure*}
	\centering
	\includegraphics[width=8.7 cm]{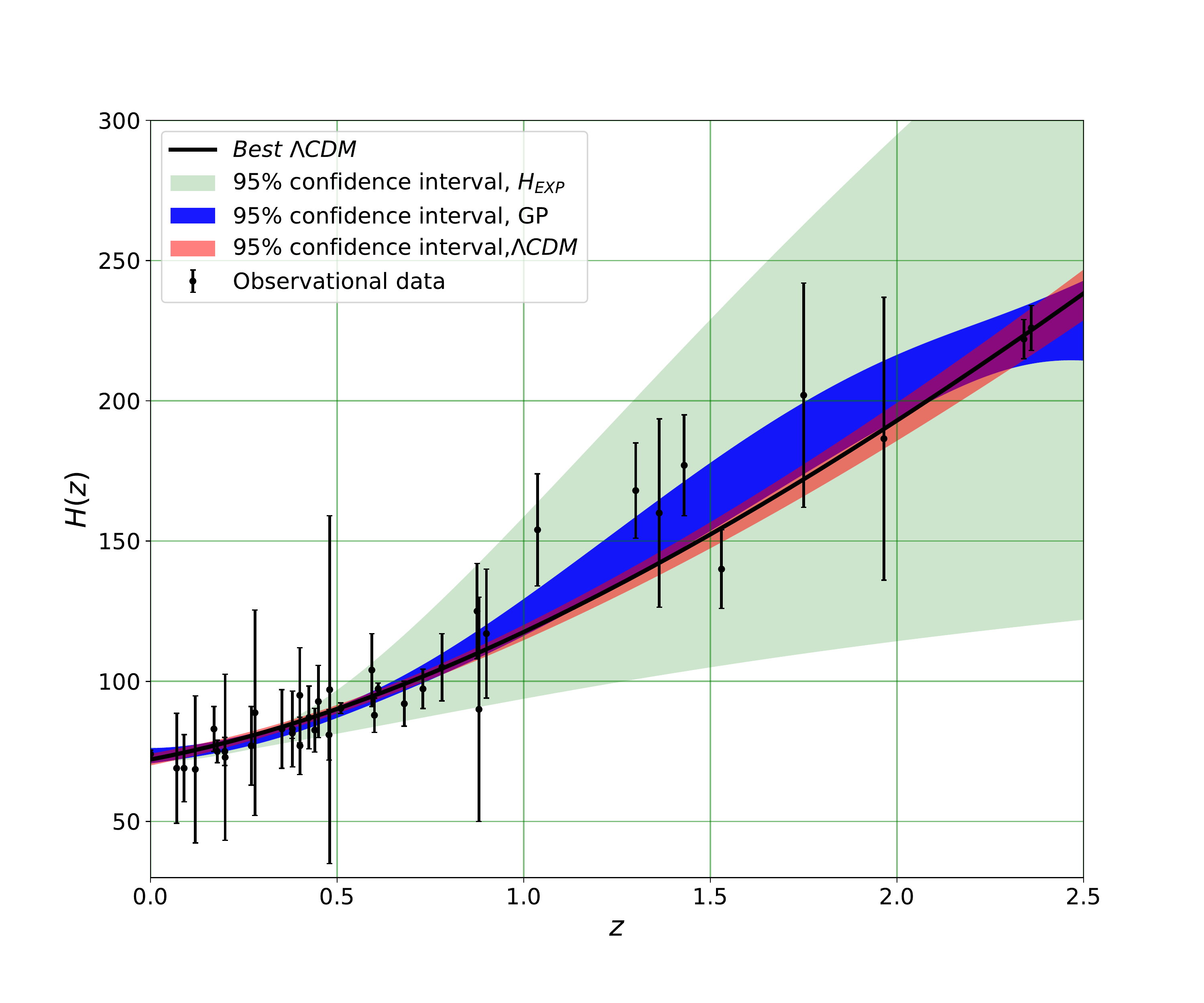}	\includegraphics[width=8.7 cm]{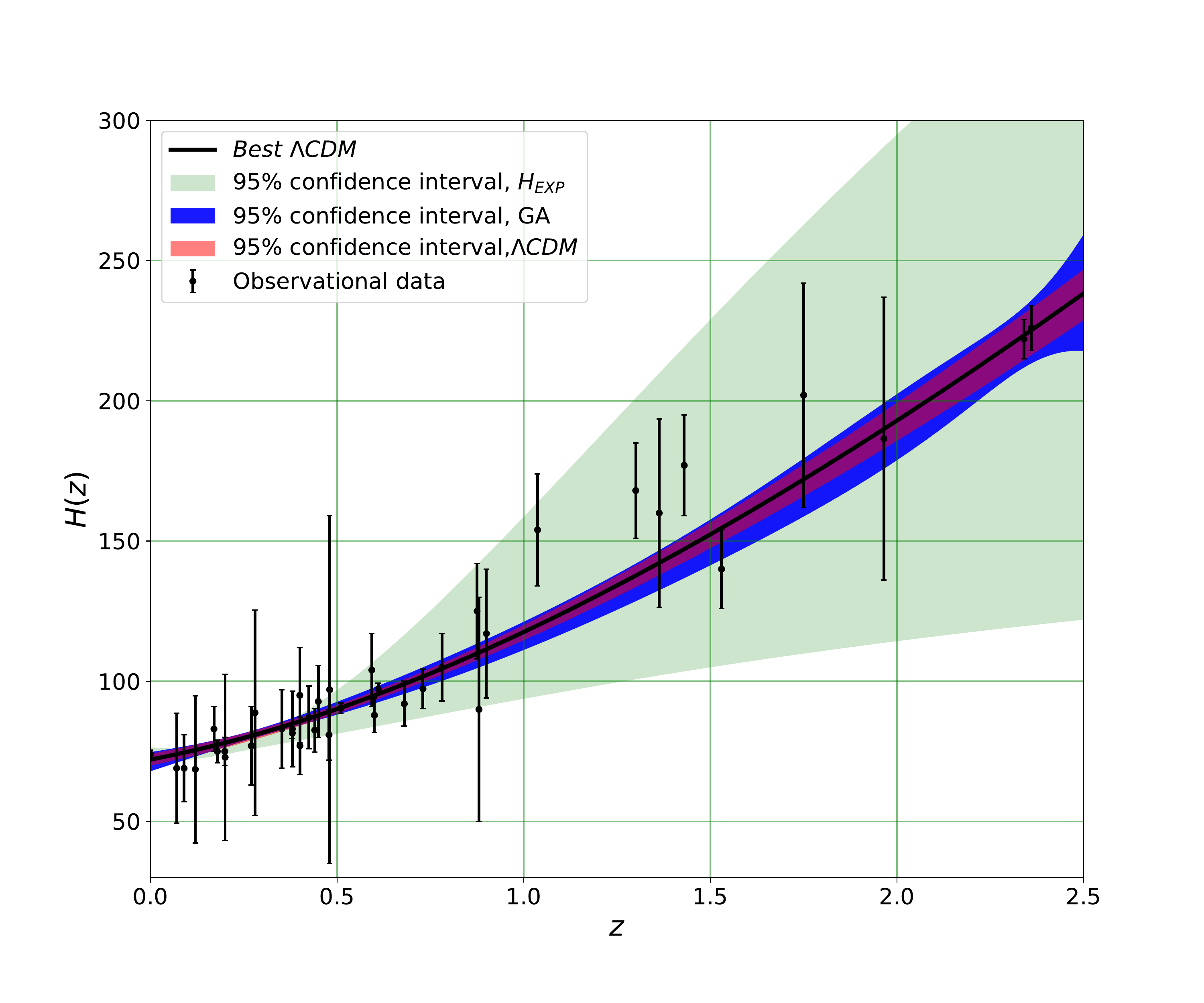}	
	\includegraphics[width=8.7 cm]{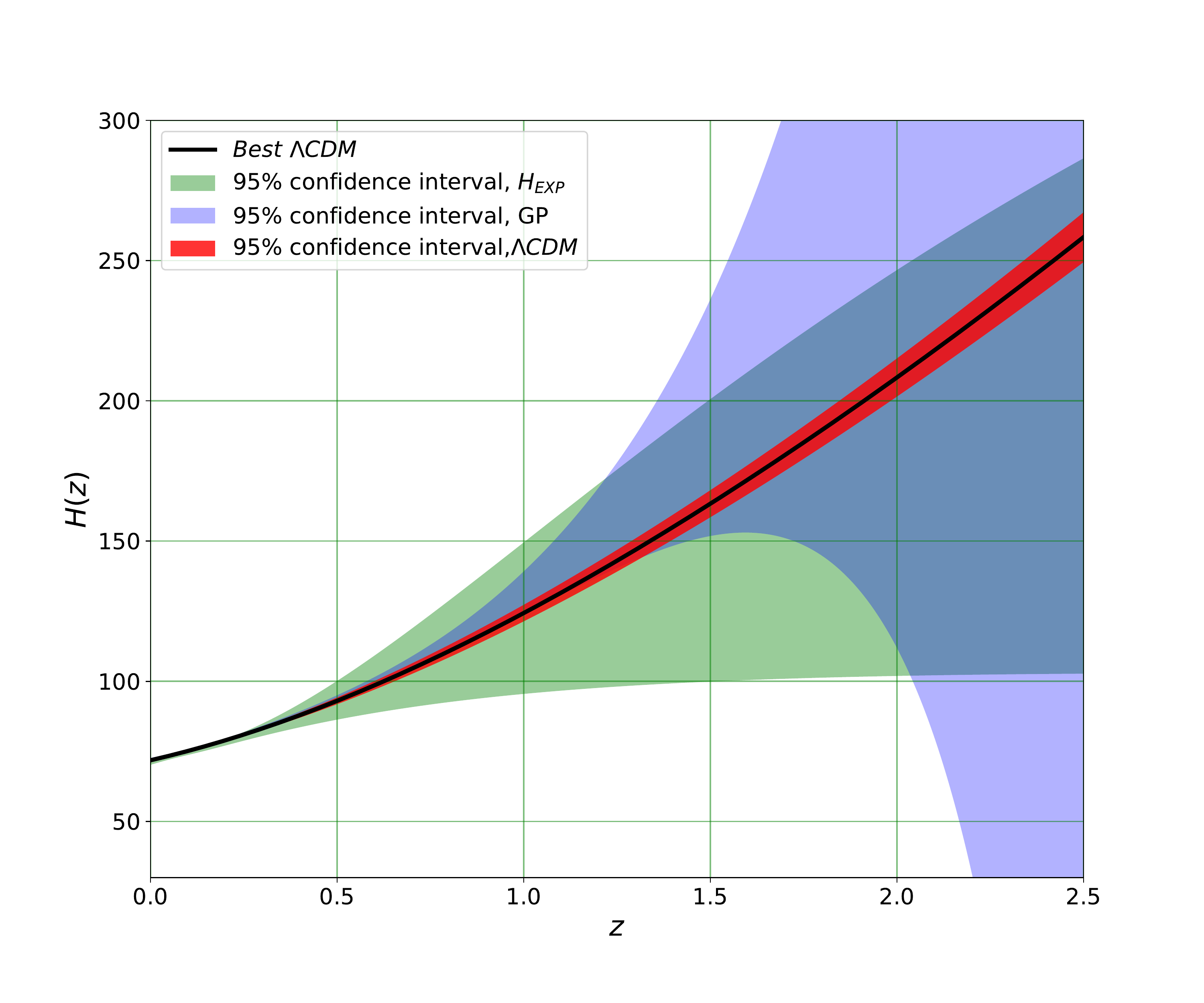} \includegraphics[width=8.7 cm]{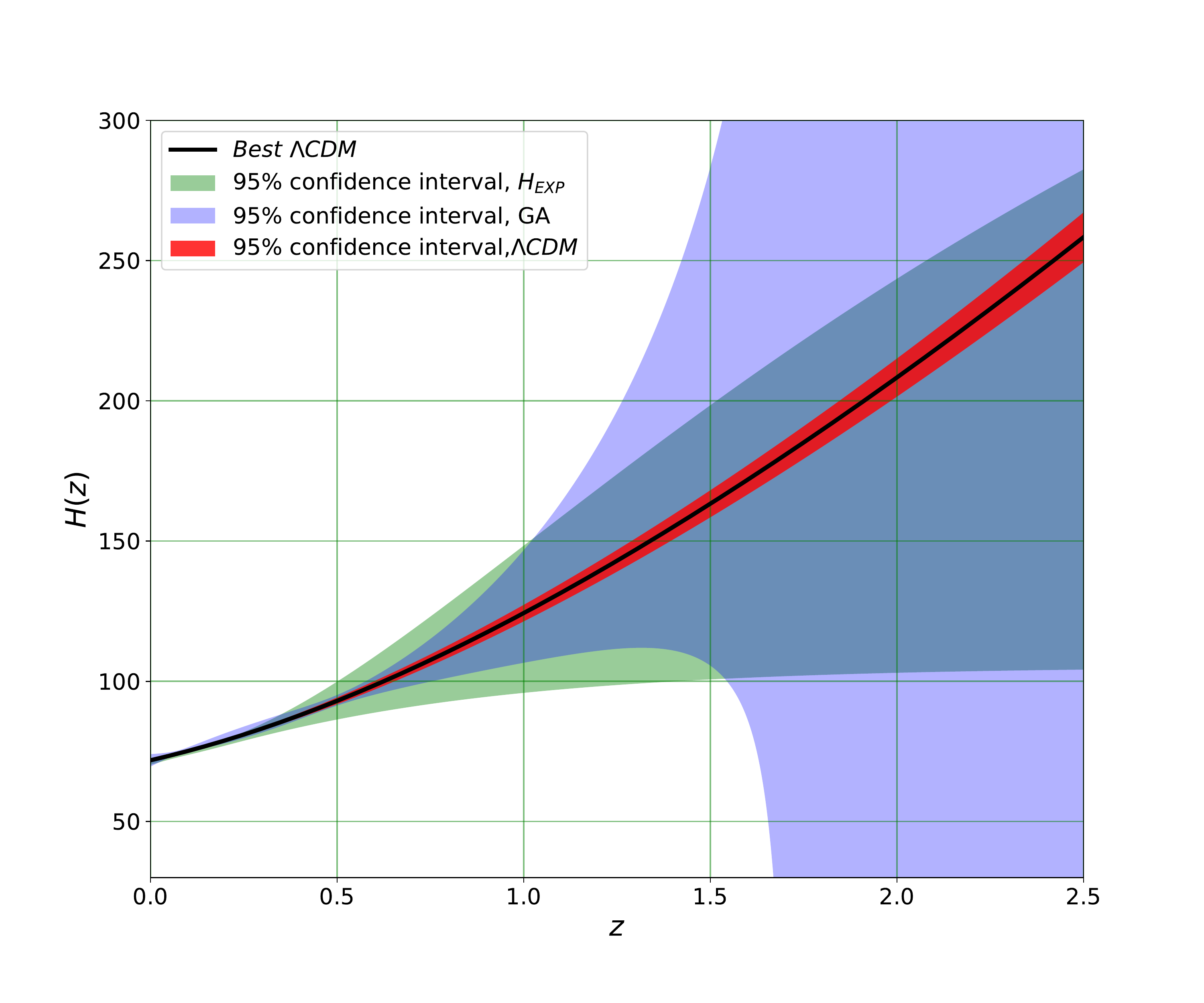}
	\caption{Upper panel: $95\%$ confidence interval of the $H(z)$ reconstructions using the Taylor expansion, $\Lambda$CDM, GP (in left panel) and GA (in right panel) as well as observational data points from Hubble data . The wide area indicate the results of the Taylor expansion. Lower panel: similar to the upper panel but in this case considering the SNIa data.}
	\label{fig:GP_H}
\end{figure*}

\begin{figure*}
	\centering
	\includegraphics[width=8.7 cm]{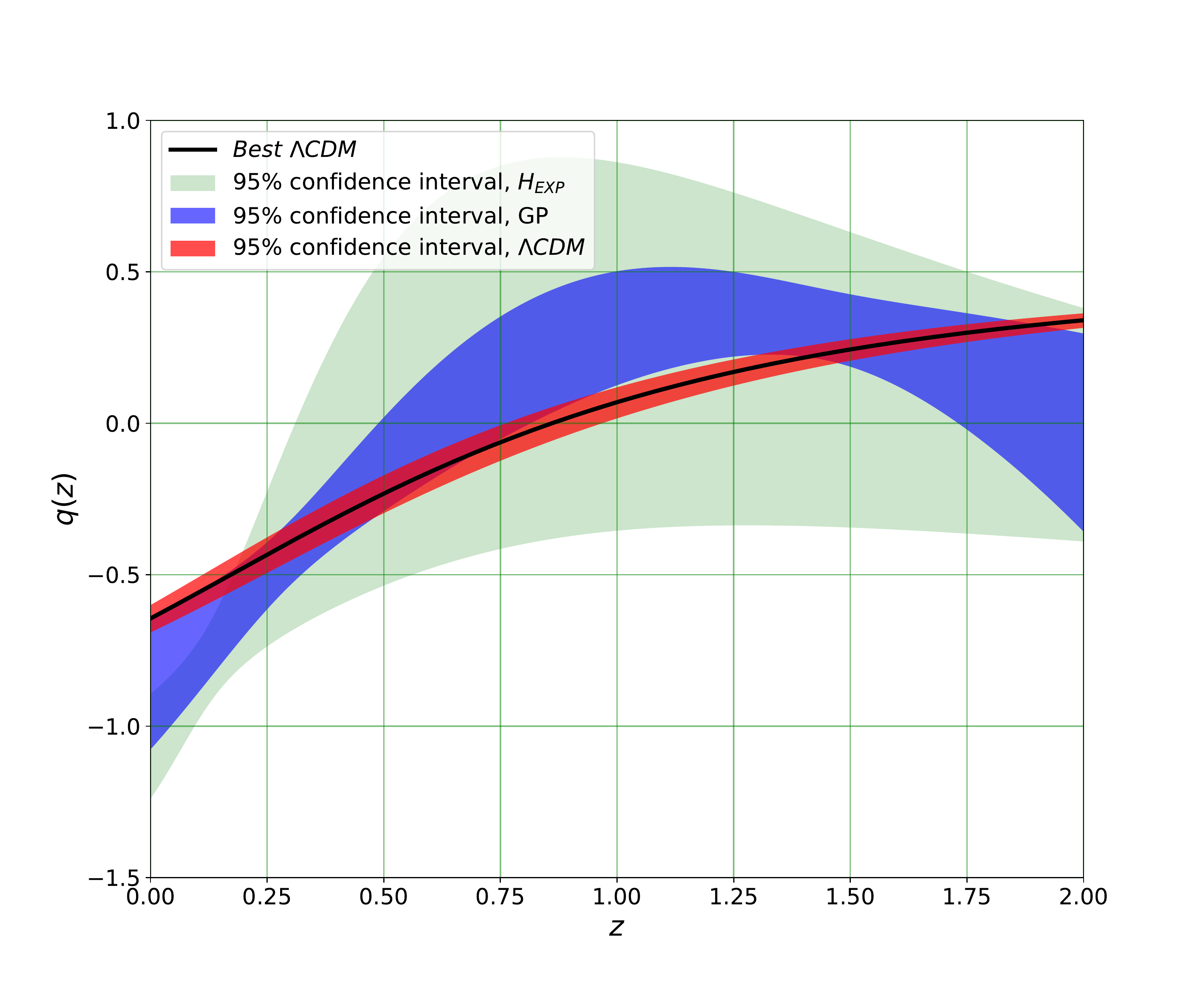}	\includegraphics[width=8.7 cm]{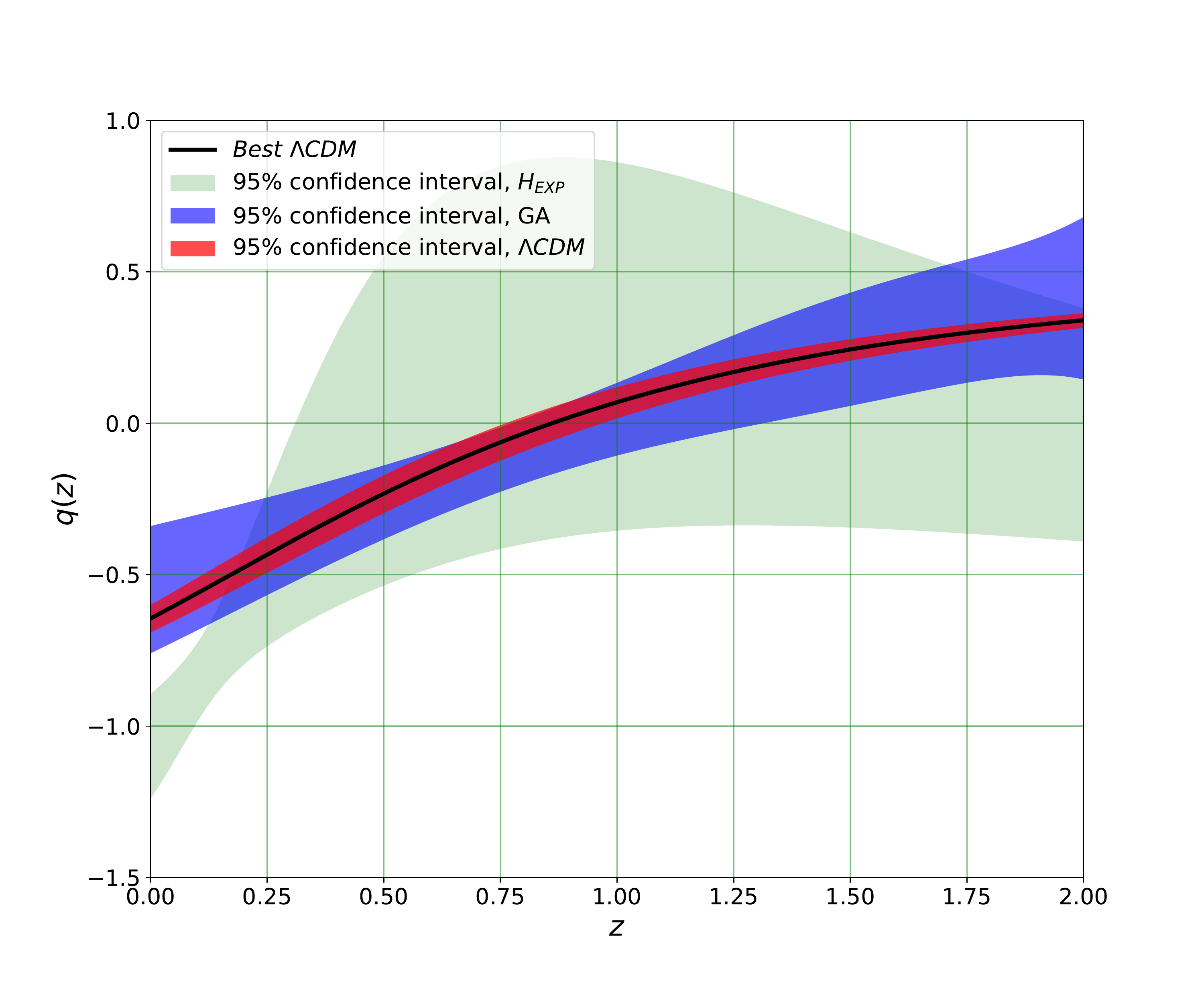}	
	\includegraphics[width=8.7 cm]{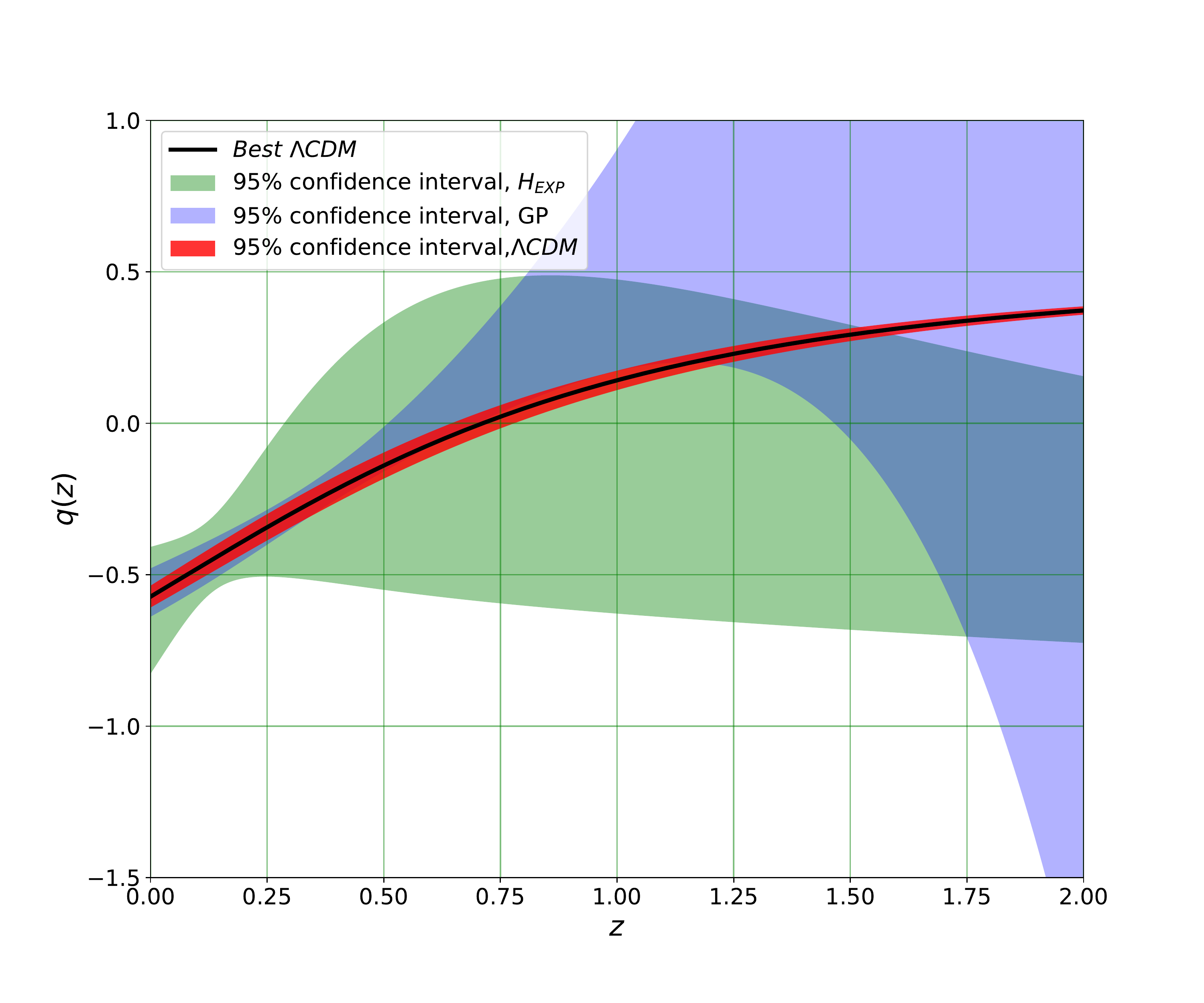} \includegraphics[width=8.7 cm]{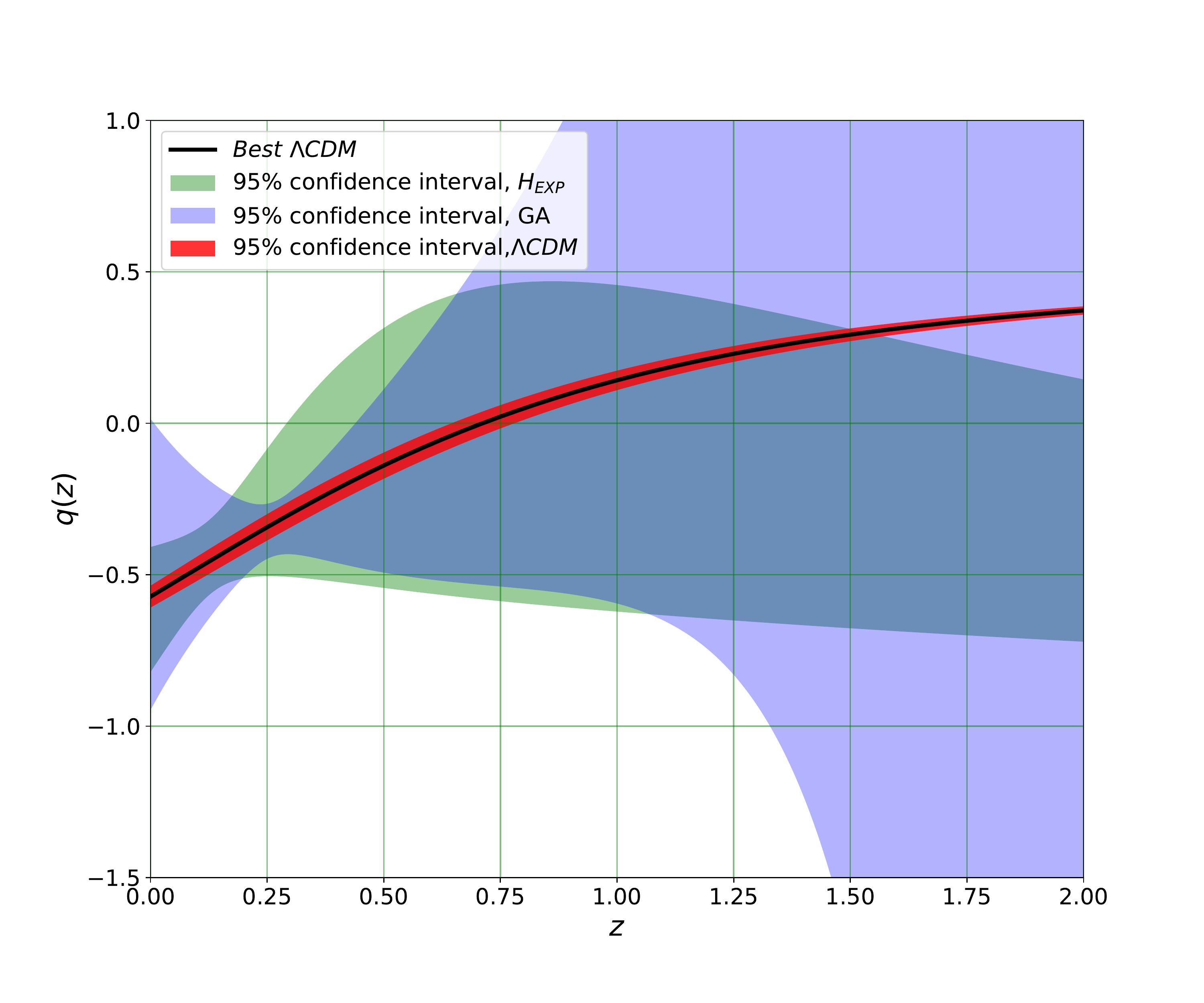}
	\caption{Upper panel: $95\%$ confidence interval of the $q(z)$ reconstructions using the Taylor expansion, $\Lambda$CDM, GP (in left panel) and GA (in right panel) from Hubble data. Lower panel: similar to the upper panel but in this case considering the SNIa data.  }
	\label{fig:GP_q}
\end{figure*}

\begin{figure*}
	\centering
	\includegraphics[width=8.5 cm]{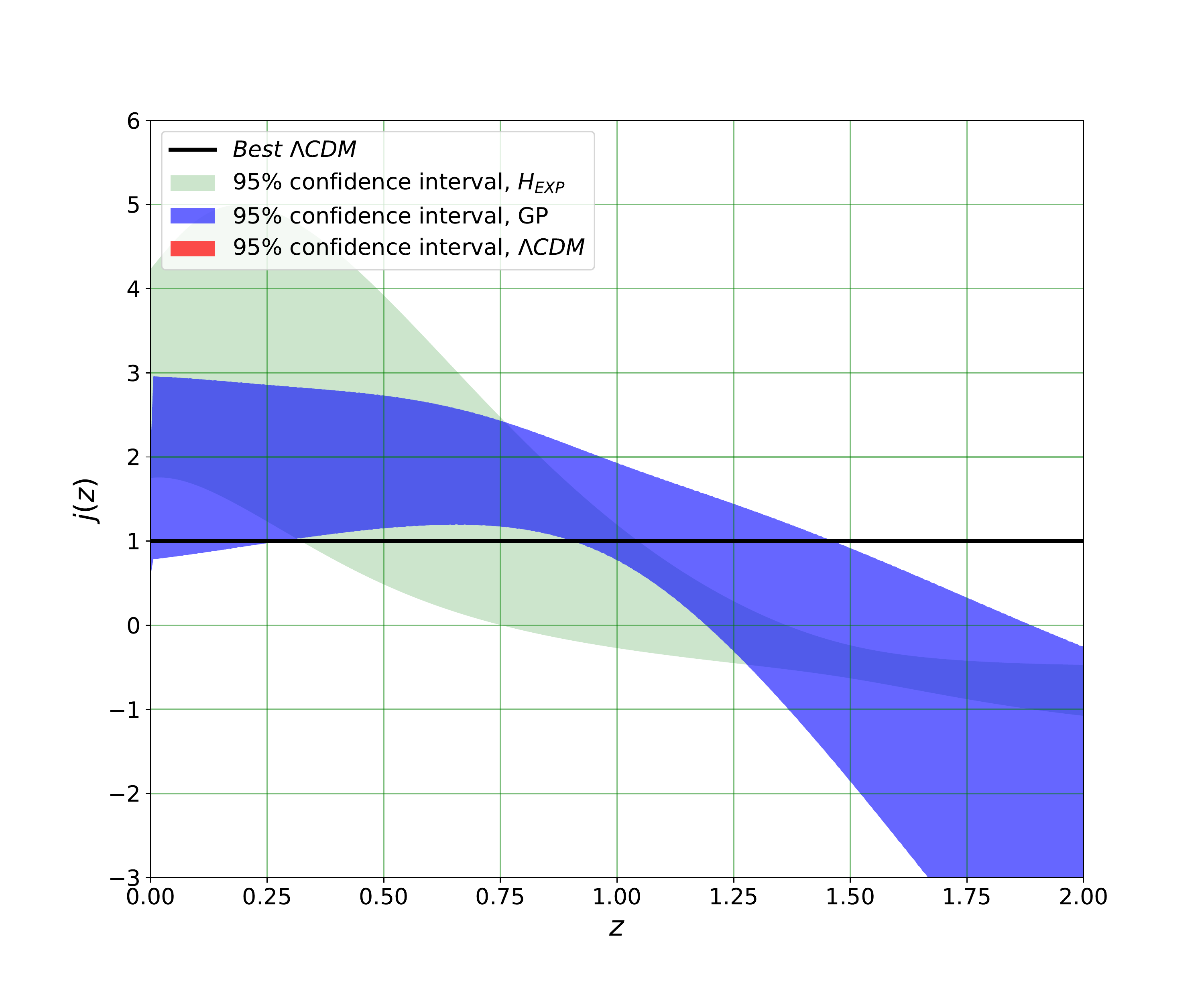}	\includegraphics[width=8.5 cm]{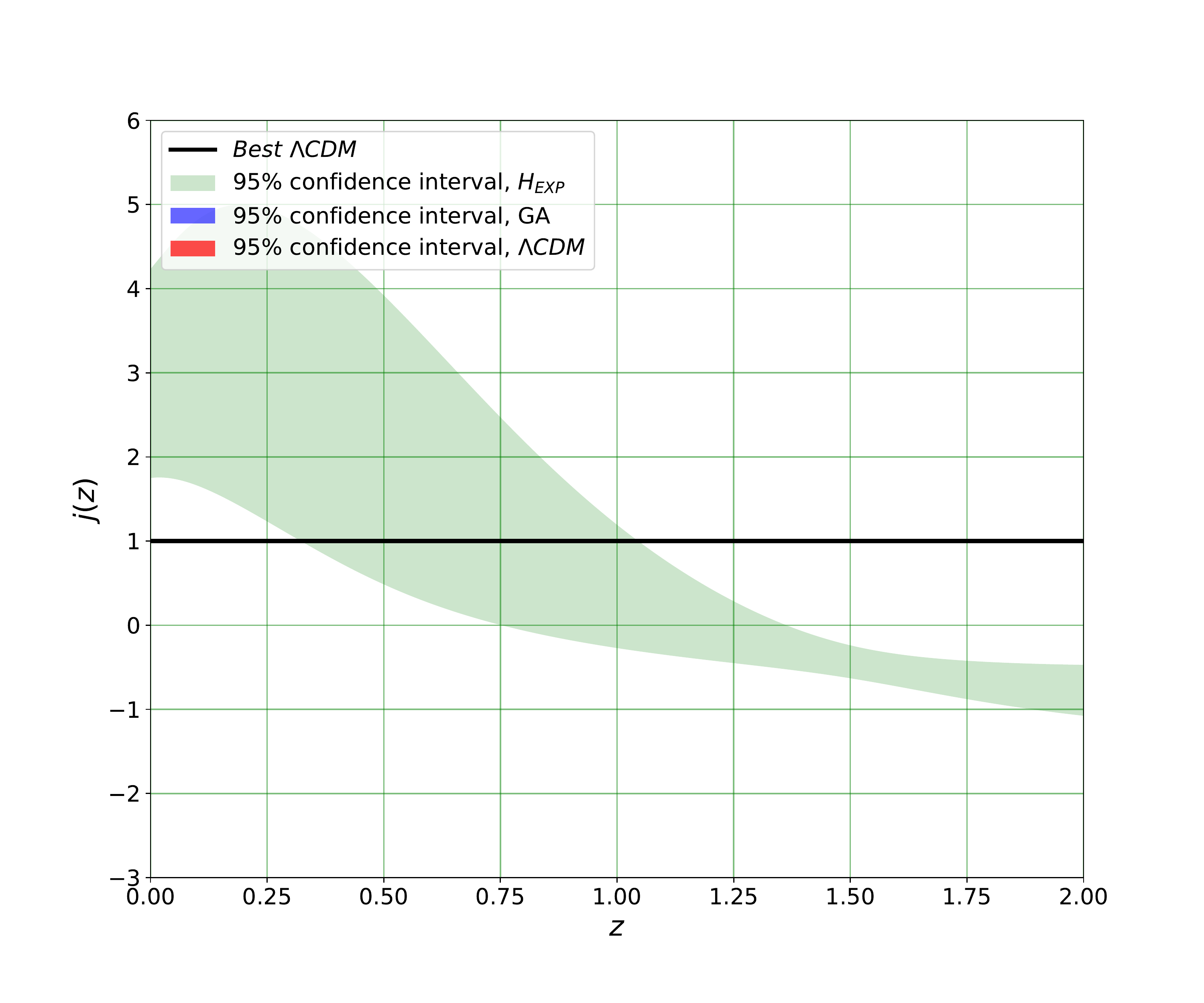}	
	\includegraphics[width=8.5 cm]{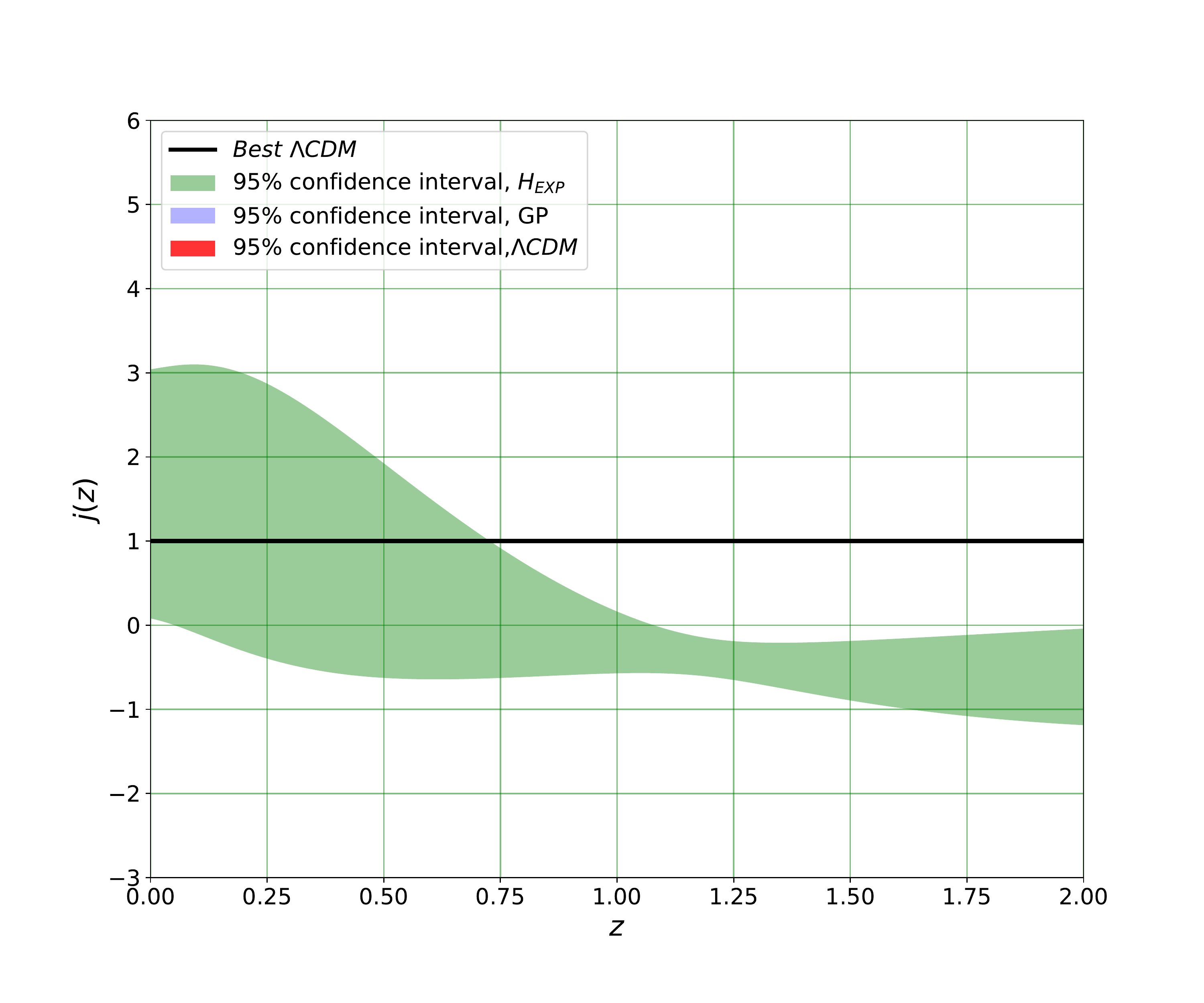} \includegraphics[width=8.5 cm]{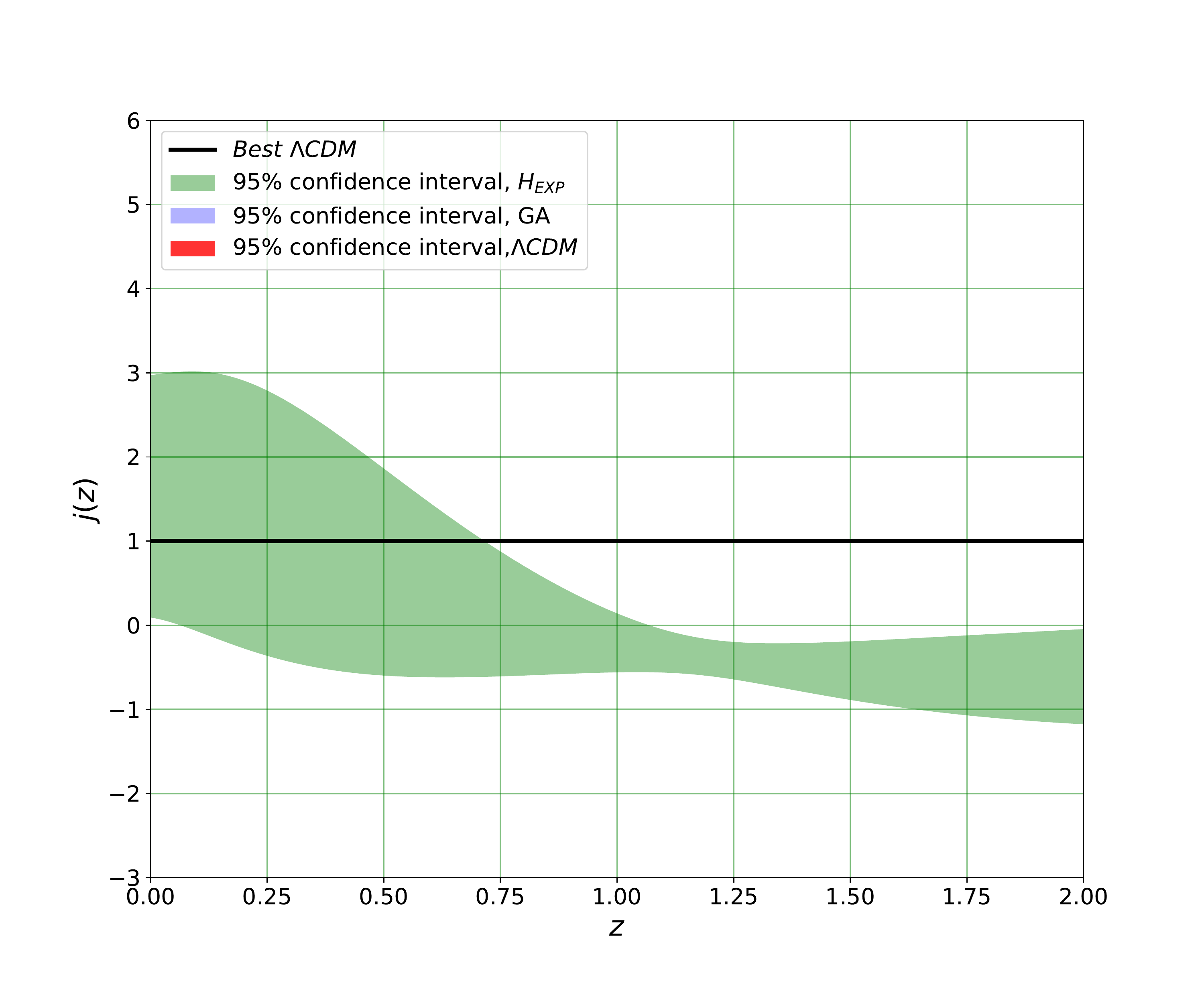}
	\caption{Upper panel: $95\%$ confidence interval of the $j(z)$ reconstructions using the Taylor expansion, $\Lambda$CDM, GP (in left panel) and GA (in right panel) from Hubble data. Lower panel: similar to the upper panel but in this case considering the SNIa data. }
	\label{fig:GP_j}
\end{figure*}

In this section, we report the main results of our analysis using $\Lambda$ cosmology and different model-independent approaches. The Hubble, deceleration and jerk parameters have been shown in Figs.(\ref{fig:GP_H}),(\ref{fig:GP_q}) and (\ref{fig:GP_j}) respectively. In all plots, the upper panels (lower panels) show results considering the Hubble data (the SNIa data). In addition,  the 95$\%$ ($1.96\sigma$) confidence interval for all approaches  as well as the best $\Lambda$CDM has been presented in these plots.

Now let summarize the main aspects of the results for different scenarios. 
\begin{itemize}
\item Reconstruction of $H(z)$ base on the Hubble data: all reconstructions and their confidence intervals are quit consistent up to redshift $z\sim0.5$. At higher redshifts, GA is still in agreement with the $\Lambda$CDM very well while GP results a small deviation at redshifts $(z\sim(1.5-2))$. In contrast to these scenarios, the cosmographic approach show a relatively higher uncertainty at this epoch. This is mainly due to the larger number of free parameters compare to the $\Lambda$CDM. The derived value of the $H_0$ in each methods has been presented in Tab.(\ref{tab:results}). 

\item Reconstruction of $H(z)$ base on the SNIa data: the $\Lambda$CDM model provides the most narrow region in this case. The results of GA and GP are consistent with each other and there is only a small upper shift in the results of GP around  $(z\sim(1.5-2))$. The interval region in cosmographic approach is similar to the Hubble case but GP and GA give a relatively larger region in particular at $z\geq 1$. This is mainly a consequence of the fact that for obtaining the Hubble parameter, we need to take a derivative of the reconstructed function in GA and GP. Notice that apart from a large uncertainty at high redshifts, all results are consistent at $1\sigma$ level.   

\item Reconstruction of $q(z)$ base on the Hubble data: The reconstructions of the deceleration parameters in this case have been shown in upper panel of the Fig.(\ref{fig:GP_q}). The result of GA is quit consistent with the $\Lambda$CDM while the GP's result deviates from the $\Lambda$CDM at some redshifts. By the way, these deviations are small and overall result is consistent. Similar to the Hubble parameter, the cosmographic approach provides a relatively larger uncertainty compare to the other two methods. Moreover, this method gives a relatively smaller value of the deceleration at present time. We think this is probably due to bias of the model and certainly should be checked in all works that only consider the cosmographic framework for a data analysis.   

\item Reconstruction of $q(z)$ base on the SNIa data: The results of all methods are consistent at present time. Uncertainties in GP and GA are quit large at higher redshifts due to the second derivative which should be computed for obtaining the deceleration. Another important point is that the uncertainty of GA is quit larger than GP at present time. This is mainly due to the uncertainty of numerical derivative at present time.(Numerical derivative usually fails at start and end points of a numerical list)  
Moreover, the results of cosmographic approach indicate a biased lower value at redshift $\sim 2$. Similar to the previous one, this is mainly due to the bias of the model.

\item Reconstruction of $j(z)$ base on the Hubble data: For the $\Lambda$CDM, the value of jerk is constant and equals to the unity. The results of GA is quit consistent with the $\Lambda$CDM but GP gives a lower value at higher redshifts. Notice that at these redshifts the uncertainty is large and so the jerk parameter doesn't deviate significantly. The most important point is the biased value of the jerk at present time in the cosmographic approach. As it is clear in the plots, it provides a value more than $3\sigma$ away from the best $\Lambda$CDM. 

\item Reconstruction of $j(z)$ base on the SNIa data: in this case, all methods give a consistent value with the $\Lambda$CDM at present time. The uncertainty in GP and GA are large at high redshifts due to the effects of taking derivatives. Similar to the deceleration, the uncertainty of the jerk at present time is quit larger in GA compare to other methods. In fact, a large uncertainty in the deceleration and taking another derivative, make the uncertainty at present time larger compared to the GP.    
\end{itemize}

Overall, our results indicate that the cosmographic approach is not actually a model-independent and acts like a model with some free parameters. In the cosmographic method, not only the uncertainties are larger than  $\Lambda$CDM due to larger number of free parameters but also the method gives some biased values considering the Hubble data. On the other hand, others two approaches provides much more consistent results. Our results suggest that it might be necessary to consider GP or GA (or both) along with the cosmographic approach to confirm the results of the method.

Finally, to present how different approach perform at present time, we compute cosmographic parameters at present time and show the results in Tab.(\ref{tab:results}). The upper (lower) panel presents results considering the Hubble data (the SNIa data) and $1\sigma (68\%)$ uncertainty of each quantity has been shown along with the best value. All $H_0$ from the Hubble data are consistent at $1\sigma$ level and the $\Lambda$CDM provide the least uncertainty. For $q_0$, the cosmographic approach provides a smaller value (more than $3\sigma$ deviation from the $\Lambda$CDM) which is due to the bias of the method. On the other hand, the results of the GA and $\Lambda$CDM are quit consistent but GP value deviates less than $2\sigma$. For the jerk parameter, we can see a clear, more than $3\sigma$, deviation in value obtained from the cosmography. In this case, the GP provides a consistent value while GA gives a value with less than $2\sigma$ deviation. Moreover, considering the SNIa data, all $H_0$ and $q_0$ are consistent with each other at $1\sigma$ level. The values of the jerk parameter are also consistent in this case but the uncertainty of the GA is relatively larger than other methods. Our results indicate that for a high quality data like the SNIa, all results are consistent and there is no discrepancy, but for the Hubble data which is less certain, the cosmographic approach might easily give a biased value. 
  
\begin{table*}
	\centering
	\begin{tabular}{|r  c  c  c  c  |}
		\hline \hline
		&        &   Hubble data    &   &   \\
		\hline
		Model$\vert$ & $H_{exp}$ & GP & GA & $\Lambda$CDM\\
		\hline 
		$H_0$$\vert$ &$73.88\pm1.34$& $73.44\pm1.40$ &$71.34\pm1.74$&$72.08\pm1.06$ \\
		\hline
		$ q_0 $$\vert$ & $-1.070\pm0.093$&$-0.856\pm0.111$&$-0.545\pm0.107$&$-0.645\pm0.023$ \\
		\hline
		$ j_0 $$\vert$  & $3.00\pm0.62$&$1.30\pm0.37$&$0.52\pm0.24$&$1.00$ \\
		\hline \hline
		&        &  Pantheon data      &   &    \\
		\hline
		Model$\vert$ & $H_{exp}$ & GP & GA & $\Lambda$CDM\\
		\hline
		$H_0$$\vert$ &$71.13\pm0.46$&$71.92\pm0.38$&$71.81\pm1.14$& $71.84\pm0.22$\\
		\hline
		$ q_0 $$\vert$ & $-0.616\pm0.105$&$-0.558\pm0.040$&$-0.466\pm0.244$&$-0.572\pm0.018$\\
		\hline
		$ j_0 $$\vert$  &$1.56\pm0.74$&$0.85\pm0.12$&$0.55\pm1.65$&$1.00$\\
		\hline \hline
		
	\end{tabular}
	\caption{A summary of the cosmographic parameters and their $1\sigma$ uncertainties which are obtained using different approaches.}\label{tab:results}
\end{table*}       

We have presented the detailed results of different model independent approach in Tab.\ref{tab:2}. In this table we report the level of consistency of the results of each method with those of $\Lambda$ cosmology as concordance model. Assuming the details of this table, one can easily compare the ability of different model independent approaches in any epochs of the universe.

\begin{table*}
	\centering
	\begin{tabular}{|r  c  c  c  c  c  c  c  c  c |}
		\hline \hline
	Function	& Method   &   & Hubble data  &   &|&   & SNIa data &    \\
		\hline
       	&    & present value & low redshift  & high redshift  &|& present value & low redshift  & high redshift   \\
		\hline
		& $H_{exp}$ & $1.7\sigma$ tension & consistent & inconsistent &|& $3.2\sigma$ tension & consistent & inconsistent  \\
$H(z)$	& GP & $1.3\sigma$ tension & consistent & consistent &|& $0.4\sigma$ tension & consistent & inconsistent  \\
	    & GA & $0.7\sigma$ tension & consistent & consistent &|& $0.1\sigma$ tension & consistent & inconsistent  \\
	    \hline
	    & $H_{exp}$ & $18.2\sigma$ tension & inconsistent & inconsistent &|& $2.4\sigma$ tension & inconsistent & inconsistent  \\
$q(z)$	& GP & $9.2\sigma$ tension & consistent & consistent &|& $0.8\sigma$ tension & consistent & inconsistent  \\
	    & GA & $4.3\sigma$ tension & consistent & consistent &|& $5.9\sigma$ tension & inconsistent & inconsistent  \\
        \hline
		& $H_{exp}$ & $3.2\sigma$ tension & inconsistent & inconsistent &|& $0.8\sigma$ tension & inconsistent & inconsistent  \\
$j(z)$	& GP & $0.8\sigma$ tension & inconsistent & inconsistent &|& $1.3\sigma$ tension & consistent & inconsistent  \\
	    & GA & $2.0\sigma$ tension & consistent & inconsistent &|& $0.3\sigma$ tension & inconsistent & inconsistent  \\		
		\hline \hline
	\end{tabular}
	\caption{The level of consistency of reconstructed cosmographic functions using  different approaches compared with the results of $\Lambda$ cosmology.}\label{tab:2}
\end{table*} 
 
\section{Conclusion}\label{sec:con}
In the presence of different cosmological models which have proposed to explain the accelerated expansion of the universe, it will be helpful if one can justify this acceleration without assuming a DE model. Different approaches have been proposed in literature for this aim, dubbed model-independent approaches.
In this paper, we have applied three different kinds of these approaches to reconstruct the cosmographic functions from the Hubble data and Pantheon sample.  Besides these approaches we consider the $\Lambda$CDM cosmology as a concordance model. The first model-independent approach is built from a Taylor series of the Hubble parameter. Although this approach is widely used in literature, it suffers from fundamental difficulties with the convergence of the series. To overcome the problem, we have used an improved redshift definition, the y-redshift $y=z/(z+1)$. For each of the data sets, we perform an MCMC algorithm to find the best fit value of the free parameters and their uncertainties. In this case, the free parameters are our cosmographic parameters, $H_0, q_0$ and $j_0$. Our results indicate that in this analysis we can not put tight constraints on the other two cosmographic parameters, $s_0$ and $l_0$ and these parameters have no impact on the fitted curve. This result is in full agreement with the results which authors of \cite{Rezaei_2020,Rezaei:2021qwd} obtained using different data sets. 

Furthermore, from the Hubble parameter, we have obtained the evolution of deceleration and jerk functions and their uncertainties as the functions of redshift. In this case, not only the uncertainties are larger than those of $\Lambda$ cosmology due to more number of the free parameters, but also  considering the Hubble data, this method gives some biased values at present time. In particular, while the value of $H_0$ is relatively consistent with the results of $\Lambda$CDM, the values of $q_0$ and $j_0$ are more than $3\sigma$ away from that of $\Lambda$ cosmology (see the results in Tab.\ref{tab:2}). Moreover, using Hubble data, the reconstructed functions obtained from $H$ expansion method have the larger uncertainties among different methods, especially at higher redshifts.

Assuming SNIa data, the results are a bit different. In this case the current value of jerk parameter is consistent with that of concordance model at about $0.8\sigma$. In this case the Hubble function reconstructed using $H$ expansion method is consistent with  $\Lambda$CDM at redshift range $z<1$, while in higher redshifts and for the other cosmographic functions, the results are disappointing.

The second model independent approach we have used in this paper is Gaussian process. The $H(z)$ and $q(z)$ functions which have been reconstructed from Hubble data using this approach are consistent with concordance model, especially at lower redshifts. Assuming SNIa data at higher redshift, the results of GP decouples from those of $\Lambda$ cosmology. In the case of jerk function, the results of GP are not acceptable, except for low redshift SNIa data points.  

Last model independent approach we have used in this work is Genetic algorithm. By GA and using Hubble data we reconstruct $H(z)$ and $q(z)$ functions, completely close to the results of $\Lambda$CDM. These results repeated when we use Hubble data to reconstruct jerk function, but just at low redshifts. Assuming SNIa data in this approach leads to good reconstruction of $H(z)$ at $z<1$. While in the other cases, GA with SNIa data failed in reconstruction of cosmographic functions.

At present time, for the Hubble data as well as the SNIa data, the predicted $H_0$ value using GA, is consistent with the $\Lambda$CDM at $<0.7\sigma$ level.
Moreover, considering the Hubble data, all of approaches give the value of $q_0$ completely away from the $q_0$ of $\Lambda$CDM, except GA which provides less deviation from the results of $\Lambda$CDM. For the jerk parameter $(j_0)$, results of GP using both of data samples is consistent  with the $\Lambda$CDM model at about $1\sigma$ confidence level. On the other hand, $H$-expansion and GA lead to good results for $j_0$, just by considering the SNIa data.
 
In addition, at intermediate redshifts $(z\sim 0.5-1.3)$, performance of the GP and GA are comparable with $\Lambda$CDM while the cosmographic approach provides a larger uncertainty. In contrast, for SNIa data at higher redshifts, the uncertainty in $q(z)$ and $j(z)$ from GP and GA, are relatively larger than those of $H$-expansion methods. This is mainly due to taking derivative of the reconstructed function. Summarizing all of the above results we can say that for reconstruction of $H(z)$, GA is the more consistent method which followed by GP. In reconstructing deceleration function, none of the approaches lead to consistent results, but among them the GP was better than the other. In order to reconstruct $j(z)$, GA and GP lead to relatively good results, while the results of $H$-expansion approach is disappointing. 
Overall, we conclude from these results that the cosmographic approach can not reconstruct cosmographic functions exactly, especially at higher redshifts. This method performs like a model with more number of the free parameters and this makes errors to be more larger than those of the $\Lambda$CDM. In addition, comparing to the other methods applied in this work, this method is not flexible enough to fit the data better than the $\Lambda$CDM.
Since assuming both of the data samples, the other two model-independent methods provide more consistent results compared to concordance model. It is confirmed that the cosmographic approach gives biased results. 
Therefore we suggest that for confirming any tension in analyzing a data set especially at high redshifts,  it might be necessary to consider a more sophisticated model-independent method like the GP and GA besides using cosmographic approach.

\bibliographystyle{aasjournal}
\bibliography{ref}
\label{lastpage}

\end{document}